\documentclass[preprint,12pt,authoryear]{elsarticle}

\usepackage[T1,T2A]{fontenc}
\usepackage[utf8]{inputenc}
\usepackage{amsmath}
\usepackage{amssymb}
\usepackage{hyperref}
\usepackage{listings}
\usepackage{graphicx}          

\newcounter{bla}

\DeclareDocumentCommand \fundiff{m}{\mathcal{D}#1}

\DeclareDocumentCommand \funint{> { \SplitList { , } } m }{\int\ProcessList{#1}{\fundiff}}

\newcommand{\Lgr}{\mathcal{L}}

\journal{Computer Physics Communications}

\begin{document}

\begin{frontmatter}



\title{Dirac: a command-line $\gamma$-matrix calculator}


\author[a]{Sergii Kutnii\corref{author}}

\begin{abstract}
A software for simplification of Dirac matrix polynomials 
that arise in particle physics problems is implemented.
\end{abstract}

\end{frontmatter}



\section{Introduction} \label{intro}

Many problems in high-energy physics require simplification of polynomials of Dirac matrices.
As an example, consider the problem of classifying possible self-interactions of relativistic fermions.
Interaction lagrangian has to be a Lorentz scalar.
Let \citep[48--54]{itzykson2012}
\begin{equation}
\begin{split}
&\Gamma_1 = 1\\
&\Gamma^\mu_2 = \gamma^\mu\\
&\Gamma^{\mu_1\mu_2}_3 = \sigma^{\mu_1\mu_2}\\
&\Gamma^{\mu}_4 = \gamma^5\gamma^\mu\\
&\Gamma_5 = \gamma^5
\end{split} \label{intro/gammabasis}
\end{equation} 
Then a generic quartic self-interaction term can be written as
\begin{equation}
\Lgr_{4Aij} = t_{A\bar{\mu}\bar{\nu}}\bar{\psi}\Gamma_i^{\bar{\mu}}\psi\bar{\psi}\Gamma_j^{\bar{\nu}}\psi, \label{intro/4-interaction}
\end{equation} 
where $\bar{\mu}, \bar{\nu}$ are Lorentz multi-indices, i.e. combinations of zero to two indices, and $t_{Aij\bar{\mu}\bar{\nu}}$ is some 
Lorentz-invariant tensor with up to four indices. 
By virtue of Weyl's theorem on invariants of orthogonal groups \cite[53]{Weyl1946}, such tensors 
can only be built from the metric and Levi-Civita symbol. If parity symmetry is required, then only metric is allowed.
Not all terms of the form \eqref{intro/4-interaction} are independent though. 
The matrices \eqref{intro/gammabasis} form a complete basis in the algebra of $4\times4$ complex matrices $gl(4,\mathbb{C})$.
It is easy to show that for the Grassmann field $\psi$
\begin{equation}
\psi\bar{\psi} = - \frac{1}{4}\bar{\psi}\psi - \frac{1}{4}\gamma_\mu\bar{\psi}\gamma^\mu\psi 
- \frac{1}{8}\sigma_{\mu\nu}\bar{\psi}\sigma^{\mu\nu}\psi 
+ \frac{1}{4}\gamma^5\gamma_\mu\bar{\psi}\gamma^5\gamma^\mu\psi
- \frac{1}{4}\gamma^5\bar{\psi}\gamma^5\psi. \label{intro/cliffordexpansion}
\end{equation}
Therefore, 
\begin{equation}
\bar{\psi}\Gamma_i^{\bar{\lambda}}\psi\bar{\psi}\Gamma_k^{\bar{\nu}}\psi = 
\Lambda_j\bar{\psi}\Gamma_i^{\bar{\lambda}}\Gamma_{j\bar{\mu}}\Gamma_k^{\bar{\nu}}\psi\bar{\psi}\Gamma_{j}^{\bar{\mu}}\psi,
\end{equation}
where summation over repeated indices is implied, and $\Lambda_j$ are coefficients in the expansion \eqref{intro/cliffordexpansion}.
But completeness also implies the decomposition
\begin{equation}
\Gamma_i^{\bar{\lambda}}\Gamma_{j\bar{\mu}}\Gamma_k^{\bar{\nu}} = 
K_{ijkl\bar{\mu}\bar{\kappa}}^{\bar{\lambda}\bar{\nu}}\Gamma_l^{\bar{\kappa}},
\end{equation}
for certain numeric coefficients $K_{ijkl\bar{\mu}\bar{\kappa}}^{\bar{\lambda}\bar{\nu}}$.
Therefore, quartic scalar combinations of fermions are subject to identities
\begin{equation}
\bar{\psi}\Gamma_i^{\bar{\kappa}}\psi\bar{\psi}\Gamma_j^{\bar{\lambda}}\psi =
{f_{ijkl}}^{\bar{\kappa}\bar{\lambda}}_{\bar{\mu}\bar{\nu}}\bar{\psi}\Gamma_k^{\bar{\mu}}\psi\bar{\psi}\Gamma_l^{\bar{\nu}}\psi. \label{intro/fierz4}
\end{equation}
These are the well-known Fierz identities \citep{Fierz1937}, \citep[161--162]{itzykson2012}. 
They can be derived manually with relative ease.

But completeness also implies an analogous set of identities for sixth-order terms
\begin{equation}
I_{ijk}^{\bar{\lambda}\bar{\mu}\bar{\nu}} = 
\bar{\psi}\Gamma_i^{\bar{\lambda}}\psi\bar{\psi}\Gamma_j^{\bar{\mu}}\psi\bar{\psi}\Gamma_k^{\bar{\nu}}\psi. \label{intro/6-terms}
\end{equation}
The need to consider such terms arose in a study of bosonization in a Nambu-Jona-Lasinio-like model with additional color symmetry \citep{kutnii2023bosonization}.
The idea of bosonization, proposed by Kikkawa \citep{Kikkawa1974}, Ebert, and Reinhardt \citep{Ebert1978}, \citep{Reinhardt1978}, is to utilize the relation
\begin{equation}
  \exp\left[\lambda{}\int\,A^2(x)d^dx\right] \sim \funint{B}\exp\left\{\int\left[\frac{1}{4\lambda}B(x)^2 + A(x)B(x)\right]d^dx\right\},
\end{equation}
a functional analog of Fourier or Laplace transform, to represent a generic four-fermion interaction term as
\begin{equation}
\begin{split}
  &\exp\left[-i\int{}G_{\bar{i}}\bar{\psi}\tau_{\bar{i}\bar{\alpha}}\psi\bar{\psi}\tau_{\bar{i}}^{\bar{\alpha}}\psi{}d^dx\right] \sim \\
  &\sim\funint{\Sigma}\exp\left\{-i\int\left[-\frac{1}{4G_{\bar{i}}}\Sigma_{\bar{i}\bar{\alpha}}\Sigma_{\bar{i}}^{\bar{\alpha}} 
  +\Sigma_{\bar{i}\bar{\alpha}}\bar{\psi}\tau_{\bar{i}}^{\bar{\alpha}}\psi \right]d^dx\right\}.
\end{split}
\end{equation}
Here
\begin{equation}
\begin{split}
&\tau_{i0}^{\bar{\alpha}} = \Gamma_i^{\bar{\alpha}}\\
&\tau_{i1}^{\bar{\alpha}} = T^a\Gamma_i^{\bar{\beta}},
\end{split} \label{intro/deftau}
\end{equation}
where $T^a$ are the generators of $su(N)$ algebra, and shorthand notation $\bar{i}$ is used for index pairs $i0, i1$.
The generating functional of the model can thus be expressed as
\begin{equation}
\begin{split}
&Z\left[\bar{\eta}\eta\right] \equiv \funint{\bar{\psi}, \psi}\exp\left\{-i\int\left[\bar{\psi}\widehat{\partial}\psi 
  + G_{\bar{i}}\bar{\psi}\tau_{\bar{i}\bar{\alpha}}\psi\bar{\psi}\tau_{\bar{i}}^{\bar{\alpha}}\psi 
  + \bar{\eta}\psi + \bar{\psi}\eta\right]d^4x\right\} \sim \\
  &\sim \funint{\bar{\psi}, \psi, \Sigma}\exp\left\{-i\int\left[\bar{\psi}\left(\widehat{\partial} + \widehat{\Sigma}\right)\psi 
  - \frac{1}{4G_{\bar{i}}}\Sigma_{\bar{i}\bar{\alpha}}\Sigma_{\bar{i}}^{\bar{\alpha}}  + \bar{\eta}\psi + \bar{\psi}\eta\right]d^4x\right\} 
\end{split}\label{intro/Bosonization}
\end{equation}
  where $\widehat{\Sigma} = \tau_{\bar{i}\bar{\alpha}}\Sigma_{\bar{i}}^{\bar{\alpha}}$.
The quartic Fierz identities \eqref{intro/fierz4} together with first fundamental theorem of invariant theory for $su(N)$ matrices \citep[21--22]{Kraft1996} 
have two major implications: that the most general massless parity-preserving NJL-like lagrangian with global $U(N)$ symmetry is
\begin{equation}
\begin{split}
\Lgr(\bar{\psi},\psi) &= \bar{\psi}i\widehat\partial\psi + G_1\bar{\psi}\psi\bar{\psi}\psi + G_2\bar{\psi}\gamma_\mu\psi\bar{\psi}\gamma^\mu\psi +\\
&+ G_3\bar{\psi}\sigma_{\mu\nu}\psi\bar{\psi}\sigma^{\mu\nu}\psi + G_4\bar{\psi}\gamma^5\gamma_\mu\psi\bar{\psi}\gamma^5\gamma^\mu\psi 
+ G_5\bar{\psi}\gamma^5\psi\bar{\psi}\gamma^5\psi, \label{intro/NJL-model}
\end{split}
\end{equation}
and that bosonization is not uniquely determined by interaction.
The lagrangian
\begin{equation}
\begin{split}
&\Lgr(\bar{\psi},\psi, \widehat{\Sigma}) = 
\bar{\psi}\left(i\widehat\partial + \widehat{\Sigma}\right)\psi 
- \frac{1}{4\lambda_{10}}\Sigma_{10}^2 - \frac{1}{4\lambda_{11}}\Sigma_{11}^a\Sigma_{11}^a -\\
&- \frac{1}{4\lambda_{20}}\Sigma_{20\,\mu}\Sigma^\mu_{20} - \frac{1}{4\lambda_{21}}\Sigma^a_{21\,\mu}\Sigma^{a\mu}_{21}
- \frac{1}{4\lambda_{30}}\Sigma_{30\,\mu\nu}\Sigma^{\mu\nu}_{30} - \frac{1}{4\lambda_{31}}\Sigma^a_{31\,\mu\nu}\Sigma^{a\mu\nu}_{31} -\\
&- \frac{1}{4\lambda_{40}}\Sigma_{40\,\mu}\Sigma^\mu_{40} - \frac{1}{4\lambda_{41}}\Sigma^a_{41\,\mu}\Sigma^{a\mu}_{41}
- \frac{1}{4\lambda_{50}}\Sigma_{50}^2 - \frac{1}{4\lambda_{51}}\Sigma^a_{51}\Sigma^a_{51}\\
\widehat{\Sigma} &=
\Sigma_{10} + T^a\Sigma^a_{11} + \gamma_\mu\Sigma^\mu_{20} + T^a\gamma_\mu\Sigma^{a\mu}_{21} +\\
&+ \sigma_{\mu\nu}\Sigma^{\mu\nu}_{30} + T^a\sigma_{\mu\nu}\Sigma^{a\mu\nu}_{31}
+ \gamma^5\gamma_\mu\Sigma^\mu_{40} + T^a\gamma^5\gamma_\mu\Sigma^{a\mu}_{41} 
+ \gamma^5\Sigma_{50} + T^a\gamma^5\Sigma^a_{51} \label{intro/NJL-bosonized}
\end{split}
\end{equation}
defines bosonization of \eqref{intro/NJL-model}, provided that
\begin{equation}
  \begin{bmatrix}
    G_1\\G_2\\G_3\\G_4\\G_5
  \end{bmatrix} = 
  \begin{bmatrix}
    \lambda_{10}\\\lambda_{20}\\\lambda_{30}\\\lambda_{40}\\\lambda_{50}
  \end{bmatrix} + 
  \begin{bmatrix}
    -\frac{N + 4}{8N}&-\frac{1}{2}&-\frac{3}{2}&\frac{1}{2}&-\frac{1}{8}\\
    -\frac{1}{2}&\frac{N - 2}{4N}&0&\frac{1}{4}&\frac{1}{8}\\
    -\frac{1}{16}&0&\frac{N - 2}{4N}&0&-\frac{1}{16}\\
    \frac{1}{8}&\frac{1}{4}&0&\frac{N - 2}{4N}&-\frac{1}{8}\\
    -\frac{1}{8}&\frac{1}{2}&-\frac{3}{2}&-\frac{1}{2}&-\frac{N + 4}{8N}
  \end{bmatrix}
  \begin{bmatrix}
    \lambda_{11}\\\lambda_{21}\\\lambda_{31}\\\lambda_{41}\\\lambda_{51}    
  \end{bmatrix}.
  \label{intro/fierz-conditions}
\end{equation}
It is clear that half of $\lambda_{\bar{i}}$ are free parameters 
and the system of equations \eqref{intro/fierz-conditions} has nontrivial solutions even if all $G_i = 0$.
This suggests a possibility to consider a generalized bosonization: to insert
\begin{equation}
\begin{split}
\text{const} &= \funint{\Sigma}\exp\left\{
-i\int\left[\frac{\lambda_{11\bar{i}}^{-\frac{1}{2}}}{2}\Sigma_{11\bar{i}\bar{\alpha}} 
- \lambda_{11\bar{i}}^\frac{1}{2}\bar{\psi}\tau_{\bar{i}\bar{\alpha}}\psi\right]\times\right.\\
 &\left.\times\left[\frac{\lambda_{11\bar{i}}^{-\frac{1}{2}}}{2}\Sigma_{11\bar{i}}^{\bar{\alpha}} 
- \lambda_{11\bar{i}}^\frac{1}{2}\bar{\psi}\tau_{\bar{i}}^{\bar{\alpha}}\psi\right]d^4x -\right.\\
 &\left.-i\int\left[
\varkappa_{21\bar{i}}^{-\frac{1}{2}}\bar{\Sigma}_{21\bar{i}\bar{\alpha}} 
- \left(\varkappa_{21\bar{i}}^\frac{1}{2}\bar{\psi}\tau_{\bar{i}\bar{\alpha}} + \lambda_{21\bar{i}}^\frac{1}{2}\bar{\psi}\tau_{\bar{i}\bar{\alpha}}\psi\bar{\psi}\right)
 \right]\times\right.\\
 &\left.\times\left[
\varkappa_{21\bar{i}}^{-\frac{1}{2}}\Sigma_{21\bar{i}}^{\bar{\alpha}} 
- \left(\varkappa_{21\bar{i}}^\frac{1}{2}\tau_{\bar{i}}^{\bar{\alpha}}\psi + \lambda_{21\bar{i}}^\frac{1}{2}\psi\bar{\psi}\tau_{\bar{i}}^{\bar{\alpha}}\psi\right)
 \right]d^4x\right\}
\end{split}
\label{intro/bosonization-gen}
\end{equation}
into path integral expression for the generating functional. This would naturally produce terms of type \eqref{intro/6-terms} 
(properties of $su(N)$ generators imply that all terms can be reduced to that form). 
Fierz identities for such terms will lead to cancellation conditions analogous to \eqref{intro/fierz-conditions}.
The main obstacle to deriving them is combinatorial explosion of complexity. 
There are, up to permutations, 35 terms of the form \eqref{intro/6-terms} (basis matrices can occur more than once within a single term), each of which has two $\psi\bar{\psi}$ factors that can be expanded along the lines of \eqref{intro/cliffordexpansion}. Thus there are 875 products of five basis matrices, each producing up to several dozen terms.

Only Lorentz-invariant sixth-order combinations of fermions can be present in the interaction lagrangian.
The even-parity ones can be built as contractions of $\psi\bar{\psi}$ with eleven Lorentz-invariant tensor products
\begin{equation}
\begin{split}
&\mathbf{\theta}_{111} = \mathbf{1}\otimes\mathbf{1}\otimes\mathbf{1}\\
&\mathbf{\theta}_{122} = \mathbf{1}\otimes\gamma_\mu\otimes\gamma^\mu\\
&\mathbf{\theta}_{133} = \mathbf{1}\otimes\sigma_{\mu\nu}\otimes\sigma^{\mu\nu}\\
&\mathbf{\theta}_{144} = \mathbf{1}\otimes\gamma^5\gamma_\mu\otimes\gamma^5\gamma^\mu\\
&\mathbf{\theta}_{155} = \mathbf{1}\otimes\gamma^5\otimes\gamma^5\\{}
&\mathbf{\theta}_{523} = \gamma^5\otimes\gamma_\mu\otimes\gamma^5\gamma^\mu\\
&\mathbf{\theta}_{223} = \gamma_\mu\otimes\gamma_\nu\otimes\sigma^{\mu\nu}\\
&\mathbf{\theta}_{443} = \gamma^5\gamma_\mu\otimes\gamma^5\gamma_\nu\otimes\sigma^{\mu\nu}\\
&\mathbf{\theta}_{243} = \epsilon_{\kappa\lambda\mu\nu}\gamma^\kappa\otimes\gamma^5\gamma^\lambda\otimes\sigma^{\mu\nu}\\
&\mathbf{\theta}_{533} = \epsilon_{\kappa\lambda\mu\nu}\gamma^5\otimes\sigma^{\kappa\lambda}\otimes\sigma^{\mu\nu}\\
&\mathbf{\theta}_{333} = \sigma_\kappa^\lambda\otimes\sigma_\lambda^\mu\otimes\sigma_\mu^\kappa
\end{split}
\end{equation}
These lead to 275 matrix products to simplify, some of them being quite complex, such as the product of five $\sigma$-matrices.
The need for an automated computation tool becomes clear. 
However, all such tools are either proprietary with prohibitive cost or their support of tensor and Dirac matrix algebra is tailored for different kinds of problems.
Cadabra \citep{KPeeters2007} could be the best existing open-source tool for the task, 
but it uses different matrix basis and is designed for computations in arbitrary number of spatio-temporal dimensions. 
The need to support arbitrary dimension numbers and arbitrary metrics makes it hard to introduce $\gamma^5$ since the latter
is an essentially four-dimensional object not easily generalizable to arbitrary dimension count. 
But simplification of Dirac matrix products required to produce Fierz identities has to take into account identities involving $\gamma^5$ too. e.g.
\begin{equation}
\gamma_\lambda\gamma_\mu\gamma_\nu = \eta_{\lambda\mu}\gamma_\nu - \eta_{\lambda\nu}\gamma_\mu + \eta_{\mu\nu}\gamma_\lambda + i\epsilon_{\lambda\mu\nu\rho}\gamma^5\gamma^\rho
\label{intro/trigamma}
\end{equation}
\citep{PBPal2007} which is valid for four-dimensional Lorentz metric only. 
Cadabra, for example, failed to produce any analog of \eqref{intro/trigamma}. 
GiNaC and sympy lack full support of the basis \eqref{intro/gammabasis} 
having no predefined construct for $\sigma_{\mu\nu}$ 
so even if they can simplify Dirac matrix polynomials, 
they would make it necessary to write some basis mapping code to present the results in the required form.
The philosophy behind \textit{dirac} is different from that of the tools mentioned above: instead of trying to solve a general problem with a general-purpose computer algebra system, 
\textit{dirac} simplifies $\gamma$-matrix polynomials in four-dimensional Minkowski spacetime 
and produces output suitable for copy-pasting into \LaTeX documents. 
The goal was to produce the most efficient solution for one highly specific use case, 
rather than competing with much older and more feature-rich computer algebra systems.

\section{Representation and algorithms} \label{repr}

Completeness of the basis \eqref{intro/gammabasis} implies that
\begin{equation}
\Gamma_i^{\bar{\alpha}}\Gamma_j^{\bar{\beta}} = {C_{ij}^{k}}^{\bar{\alpha}\bar{\beta}}_{\bar{\gamma}}\Gamma^{\bar{\gamma}}_k.
\label{repr/defconstants} 
\end{equation}
The multiplication structure constants ${C_{ij}^{k}}^{\bar{\alpha}\bar{\beta}}_{\bar{\gamma}}$ can be grouped into 
\textit{pseudo-matrices}.
Let 
\begin{equation}
\mathbf{\Gamma}^{\bar{\mu}} = 
\begin{bmatrix}
1&\gamma^\mu & \sigma^{\mu_1\mu_2} & \gamma^5\gamma^\mu & \gamma^5 \label{repr/basisrow}
\end{bmatrix}
\end{equation}
Then \eqref{repr/defconstants} is equivalent to
\begin{equation}
\Gamma_i^{\bar{\mu}}\mathbf{\Gamma}^{\bar{\nu}} = \mathbf{\Gamma}^{\bar{\lambda}}
\left(\mathbf{C}_i^{\bar{\mu}}\right)_{\bar{\lambda}}^{\bar{\nu}}.
\end{equation}
The nice thing about pseudo-matrices $\mathbf{C_i^{\bar{\mu}}}$ is that they form a representation
of $\Gamma_i^{\bar{\mu}}$: it is easy to verify that multiplication of $\Gamma$ corresponds to multiplication of the respective pseudo-matrices $\mathbf{C}$ in the same order plus contraction of matching tensor indices. 
Thus, all pseudo-matrices can be constructed recursively from basic multiplicative identities,
a good reference for which can be found in \citep{PBPal2007}. Let the metric be denoted with $\eta$, and imaginary unit with $I$. 
The pseudo-matrix counterparts to $\gamma^\mu$ are
\begin{equation}
\begin{split}
&\mathbf{C}_2^\mu = 
\begin{bmatrix}
0 & \eta^{\mu\nu} & 0 & 0 & 0 \\
\delta^\mu_\lambda & 0 & I\left(\eta^{\mu\nu_1}\delta_\lambda^{\nu_2} - \eta^{\mu\nu_2}\delta_\lambda^{\nu_1}\right) & 0 & 0 \\
0 & -\frac{I}{2}\left(\delta_{\lambda_1}^\mu\delta_{\lambda_2}^\nu - \delta_{\lambda_2}^\mu\delta_{\lambda_1}^\nu\right) & 0 & \frac{1}{2}{\epsilon^{\mu\nu}}_{\lambda_1\lambda_2} & 0 \\
0 & 0 & -{\epsilon^{\mu\nu_1\nu_2}}_\lambda & 0 & -\delta^\mu_\lambda \\
0 & 0 & 0 & -\eta^{\mu\nu} & 0 
\end{bmatrix}
\end{split}
\end{equation}
and $\gamma^5$ is represented with
\begin{equation}
\begin{split}
&\mathbf{C}_5 = 
\begin{bmatrix}
0 & 0 & 0 & 0 & 1\\
0 & 0 & 0 & \delta_\lambda^\nu & 0\\
0 & 0 & -\frac{I}{2}{\epsilon^{\nu_1\nu_2}}_{\lambda_1\lambda_2} & 0 & 0\\
0 & \delta_\lambda^\nu & 0 & 0 & 0\\
1 & 0 & 0 & 0 & 0
\end{bmatrix}
\end{split}
\end{equation}

Given that, one can compute
\begin{equation}
\left(\mathbf{C}_4^\mu\right)^{\bar{\nu}}_{\bar{\lambda}} = 
\left(\mathbf{C}_5\right)^{\bar{\kappa}}_{\bar{\lambda}}\left(\mathbf{C}_2^\mu\right)_{\bar{\kappa}}^{\bar{\nu}} = 
\resizebox{0.5\textwidth}{!}
     {$%
\begin{bmatrix}
0 & 0 & 0 & -\eta^{\mu\nu} &0 \\
0 & 0 & -{\epsilon^{\mu\nu_1\nu_2}}_\lambda & 0 & -\delta^\mu_\lambda \\
0 & -\frac{1}{2}{\epsilon^{\nu_1\nu_2}}_{\lambda_1\lambda_2} & 0 & 
-\frac{I}{2}\left(\delta_{\lambda_1}^\mu\delta_{\lambda_2}^\nu - \delta_{\lambda_2}^\mu\delta_{\lambda_1}^\nu\right) & 0\\
\delta^{\mu}_\lambda & 0 & I\left(\eta^{\mu\nu_1}\delta_\lambda^{\nu_2} - \eta^{\mu\nu_2}\delta_\lambda^{\nu_1}\right)
& 0 & 0\\
0 & \eta^{\mu\nu} & 0 & 0 & 0
\end{bmatrix}
$}
\end{equation}
\normalsize
and the most complicated of all
\begin{equation}
\begin{split}
&\left(\mathbf{C}_3^{\mu_1\mu_2}\right)^{\bar{\nu}}_{\bar{\lambda}} = 
\frac{I}{2}\left[
\left(\mathbf{C}_2^{\mu_1}\right)_{\bar{\lambda}}^{\bar{\kappa}}
\left(\mathbf{C}_2^{\mu_2}\right)_{\bar{\kappa}}^{\bar{\nu}} - 
\left(\mathbf{C}_2^{\mu_2}\right)_{\bar{\lambda}}^{\bar{\kappa}}
\left(\mathbf{C}_2^{\mu_1}\right)_{\bar{\kappa}}^{\bar{\nu}}
\right] = \\
&=
\resizebox{0.8\textwidth}{!}
     {$%
\begin{bmatrix}
0&0&\eta^{\mu_1\nu_1}\eta^{\mu_2\nu_2} - \eta^{\mu_2\nu_1}\eta^{\mu_1\nu_2}&0&0\\
0&I\left(\delta^{\mu_1}_{\lambda}\eta^{\mu_2\nu} - \delta^{\mu_2}_\lambda\eta^{\mu_1\nu}\right)&
0&-{\epsilon^{\mu_1\mu_2\nu}}_\lambda&0\\
\frac{1}{2}\left(\delta^{\mu_1}_{\lambda_1}\delta^{\mu_2}_{\lambda_2} - \delta^{\mu_2}_{\lambda_1}\delta^{\mu_1}_{\lambda_2}\right)&
0
&\begin{array}{l}
\frac{I}{2}\left[\delta^{\mu_1}_{\lambda_1}\delta^{\nu_2}_{\lambda_2}\eta^{\mu_2\nu_1} 
-\delta^{\mu_2}_{\lambda_1}\delta^{\nu_2}_{\lambda_2}\eta^{\mu_1\nu_1} -\right.\\
-\delta^{\mu_1}_{\lambda_1}\delta^{\nu_1}_{\lambda_2}\eta^{\mu_2\nu_2}
+\delta^{\mu_2}_{\lambda_1}\delta^{\nu_1}_{\lambda_2}\eta^{\mu_1\nu_2} -\\
-\delta^{\mu_1}_{\lambda_2}\delta^{\nu_2}_{\lambda_1}\eta^{\mu_2\nu_1} +
\delta^{\mu_2}_{\lambda_2}\delta^{\nu_2}_{\lambda_1}\eta^{\mu_1\nu_1} +\\
\left.+\delta^{\mu_1}_{\lambda_2}\delta^{\nu_1}_{\lambda_1}\eta^{\mu_2\nu_2}
-\delta^{\mu_2}_{\lambda_2}\delta^{\nu_1}_{\lambda_1}\eta^{\mu_1\nu_2}\right]
\end{array}
&0&-\frac{I}{2}{\epsilon^{\mu_1\mu_2}}_{\lambda_1\lambda_2}\\
0&-{\epsilon^{\mu_1\mu_2\nu}}_\lambda&0&
I\left(\eta^{\mu_2\nu}\delta^{\mu_1}_\lambda - \eta^{\mu_1\nu}\delta^{\mu_2}_\lambda\right)&0\\
0&0&-I\epsilon^{\mu_1\mu_2\nu_1\nu_2}&0&0
\end{bmatrix}
$}
\end{split}
\end{equation}

Then any product of $\Gamma$-matrices can be represented as follows:
\begin{equation}
\Gamma_{i_1}^{\bar{\mu}_1}
\ldots\Gamma_{i_k}^{\bar{\mu}_k}
\ldots\Gamma_{i_n}^{\bar{\mu}_n} \rightarrow
{\left(\mathbf{C}_{i_1}^{\mu_1}\right)}^{\bar{\lambda}_2}_{\bar{\lambda}_1}
\ldots{\left(\mathbf{C}_{i_k}^{\mu_k}\right)}_{\bar{\lambda}_k}^{\bar{\lambda}_{k+1}}
\ldots{\left(\mathbf{C}_{i_n}^{\mu_n}\right)}_{\bar{\lambda}_n}^{\bar{\lambda}_{n+1}}.
\label{repr/productrepr}
\end{equation}

On the other hand, any matrix in $gl(4, \mathbb{C})$ can be written as product of 
the basis row \eqref{repr/basisrow} and a coefficient column vector. In particular,
\begin{equation}
1 \rightarrow \begin{bmatrix}1\\0\\0\\0\\0\end{bmatrix}. \label{repr/unit-vector}
\end{equation}
Any matrix can be multiplied by $1$. 
This implies that column vector representation of the left hand side in \eqref{repr/productrepr}
can be obtained by multiplying its pseudo-matrix counterpart 
with the right hand side of \eqref{repr/unit-vector} on the right,
which is equivalent to taking the first column of the rightmost pseudo-matrix
and multiplying it by the remaining pseudo-matrices on the left in the right to left order.

All that remains is to simplify the components of the resulting column vector
which are polynomials in the metric, Kronecker, and Levi-Civita symbols with complex coefficients.
This can be done in three stages. First, Levi-Civita powers can be expanded using
\begin{equation}
\epsilon_{\mu_1\mu_2\mu_3\mu_4}\epsilon_{\nu_1\nu_2\nu_3\nu_4} = 
-\sum_{P}\text{sgn}(P)\eta_{\mu_1P\nu_1}\eta_{\mu_2P\nu_2}\eta_{\mu_3P\nu_3}\eta_{\mu_4P\nu_4},
\end{equation}
where $P$ labels all permutations of $\nu_1,\ldots,\nu_4$: 
\begin{equation}
P\left\{\mu_1, \mu_2, \mu_3, \mu_4\right\} = \left\{P\mu_1, P\mu_2, P\mu_3, P\mu_4\right\},
\end{equation}
so if $P\left\{\mu_1, \mu_2, \mu_3, \mu_4\right\} = \left\{\mu_2, \mu_3, \mu_4, \mu_1\right\}$ then
$P\mu_1 = \mu_2, P\mu_2 = \mu_3$ and so on.
Then all contractible indices can be contracted. 
using the fact that contraction of any tensor with the metric or Kronecker
amounts to simple index replacement, and Levi-Civita symbol with any index pair contracted is zero.
Contraction guarantees that all remaining indices are free. 
Then all terms having same tensor composition up to permutations of indices in constituent tensors
can be collected, taking into account Levi-Civita index permutation signs.

The only remaining major part of a $\Gamma$-matrix calculator is expression parser. 
A version of shunting yard algorithm was implemented. Input syntax is described in the next section.  

\section{The software} \label{app}
The \textit{dirac} software is available on github \citep{DiracGithub}.
It is implemented in C++20 and uses CMake (min version 3) as its build system. 
Ruby 3 language interpreter, ``open3'' Ruby gem, and \LaTeX{} installation with \verb|pdflatex| are required to run example scripts.
The standard CMake build procedure (from the project root)
\begin{lstlisting}[language=bash]
mkdir build
cd build
cmake ..
make
\end{lstlisting}
produces ``dirac'' executable in the ``build'' folder.

There are two main ways to invoke the executable. The first is to provide the expression to simplify with \verb|-e| key.
In this case the executable simplifies the expression, prints the result to the console and exits.
Example:
\begin{lstlisting}[breaklines,language=tex]
./dirac -e "\gamma_\mu\gamma_\nu"
\eta_{\nu\mu} - I\eta_{\mu\omega_{1}}\eta_{\nu\omega_{2}}\sigma^{\omega_{1}\omega_{2}}
\end{lstlisting}
(the quotes around the expression prevent the terminal from eating the backslashes).
This mode is mainly useful for scripting. 
The ``examples'' folder contains Ruby scripts that demonstrate how scripting can be used 
to batch computations and generate \LaTeX{} output from the results.

There are other command line keys as well. They affect  input processing and output formatting.

If no expression is provided via command line, the application starts an interactive shell.
The shell accepts three primary types of input.

\textbf{Quit-expression} is the single word \verb|quit| which exits the shell.

\textbf{Set-expression}:
\begin{lstlisting}
dirac:> #set <var-name> <var-value>
\end{lstlisting}
sets the variable name identified by \verb|var-name| to \verb|var-value|.
The variables are documented below.

\textbf{mode}: arithmetic mode. Possible values: \verb|float| and \verb|rational|. Default is \verb|rational|.
Equivalent command line option \verb|-m|. Example:
\begin{lstlisting}[breaklines,language=tex]
dirac:> #set mode float
dirac:> \gamma5\sigma^{\mu\nu}
-0.500000I{\epsilon^{\mu\nu}}_{\omega_{1}\omega_{2}}\sigma^{\omega_{1}\omega_{2}}
dirac:> #set mode rational
dirac:> \gamma5\sigma^{\mu\nu}
 - \frac{I}{2}{\epsilon^{\mu\nu}}_{\omega_{1}\omega_{2}}\sigma^{\omega_{1}\omega_{2}}
\end{lstlisting}
or
\begin{lstlisting}
./dirac -m float
\end{lstlisting}

The \verb|mode| variable also affects input parsing. Floating point numeric values are acceptable in float mode 
and are considered errors when the mode is rational.

The rational arithmetic implementation is the simplest possible. 
A rational number is implemented as a pair of two largest system integer types.
Since
\begin{equation}
\frac{a}{b} + \frac{c}{d} = \frac{ad + bc}{bd},
\end{equation}
longer rational expressions will tend to result in larger numerators and denominators.
Thus, the implementation is prone to integer overflows in very large $\gamma$-matrix expressions, 
when the result of an operation on integers is larger than the largest integer representable by the machine. 
It is not clear, however, whether such overflows will happen in any practical example.
Therefore a conscious decision was made to retain the naive rational implementation 
to avoid either adding library dependencies or implementing arbitrary precision arithmetic from scratch. 
A more robust implementation may be added in the future if there is user demand for it.

\textbf{line\_terms}: number of terms per output line. Possible values: integers or \verb|inf| (meaning ``infinity''). 
Default: \verb|inf|.
When this variable is set to a nonzero integer constant, \LaTeX{} line breaks (as if inside `split' environment) are inserted
after each \verb|line_terms| terms. Example:
\begin{lstlisting}[language=tex,breaklines]
dirac:> #set line_terms 2
dirac:> \gamma_\kappa\gamma_\lambda\gamma_\mu\gamma_\nu
&\eta_{\lambda\kappa}\eta_{\nu\mu} + \eta_{\mu\lambda}\eta_{\nu\kappa}  -\\
&-\eta_{\mu\kappa}\eta_{\nu\lambda} + \left[ - I\eta_{\kappa\omega_{1}}\eta_{\lambda\omega_{2}}\eta_{\nu\mu}  -\right.\\
&\left. - I\eta_{\kappa\omega_{1}}\eta_{\mu\lambda}\eta_{\nu\omega_{2}} + I\eta_{\kappa\omega_{1}}\eta_{\mu\omega_{2}}\eta_{\nu\lambda} + \right.\\
&\left.+I\eta_{\kappa\lambda}\eta_{\omega_{1}\nu}\eta_{\omega_{2}\mu}   - I\eta_{\kappa\mu}\eta_{\omega_{1}\nu}\eta_{\omega_{2}\lambda} + \right.\\
&\left.+I\eta_{\kappa\nu}\eta_{\omega_{1}\mu}\eta_{\omega_{2}\lambda}\right]\sigma^{\omega_{1}\omega_{2}}   - I\epsilon_{\lambda\mu\nu\kappa}\gamma^5
dirac:> #set line_terms inf
dirac:> \gamma_\kappa\gamma_\lambda\gamma_\mu\gamma_\nu
\eta_{\lambda\kappa}\eta_{\nu\mu} + \eta_{\mu\lambda}\eta_{\nu\kappa}  -\eta_{\mu\kappa}\eta_{\nu\lambda} + \left[ - I\eta_{\kappa\omega_{1}}\eta_{\lambda\omega_{2}}\eta_{\nu\mu}   - I\eta_{\kappa\omega_{1}}\eta_{\mu\lambda}\eta_{\nu\omega_{2}} + I\eta_{\kappa\omega_{1}}\eta_{\mu\omega_{2}}\eta_{\nu\lambda} + I\eta_{\kappa\lambda}\eta_{\omega_{1}\nu}\eta_{\omega_{2}\mu}   - I\eta_{\kappa\mu}\eta_{\omega_{1}\nu}\eta_{\omega_{2}\lambda} + I\eta_{\kappa\nu}\eta_{\omega_{1}\mu}\eta_{\omega_{2}\lambda}\right]\sigma^{\omega_{1}\omega_{2}}   - I\epsilon_{\lambda\mu\nu\kappa}\gamma^5
\end{lstlisting}
Equivalent command line option: \verb|-l|:
\begin{lstlisting}
./dirac -l 2
\end{lstlisting}

\textbf{dummy}: dummy index template. Possible values: string literals. 
Greek letters in \LaTeX{} notation are good choice.
Default is \verb|\omega|.
Equivalent command line option: \verb|-d|. Example:
\begin{lstlisting}[breaklines,language=tex]
dirac:> \gamma_\mu\gamma_\nu
\eta_{\nu\mu}   - I\eta_{\mu\omega_{1}}\eta_{\nu\omega_{2}}\sigma^{\omega_{1}\omega_{2}}
dirac:> #set dummy \sigma
dirac:> \gamma_\mu\gamma_\nu
\eta_{\nu\mu}   - I\eta_{\mu\sigma_{1}}\eta_{\nu\sigma_{2}}\sigma^{\sigma_{1}\sigma_{2}}
\end{lstlisting}
or
\begin{lstlisting}[breaklines,language=tex]
./dirac -d "\sigma"
\end{lstlisting}

\textbf{apply\_symmetry}: controls whether the coefficient terms at $\sigma^{\mu\nu}$ are merged in the output by taking into account $\sigma$'s antisymmetry.
Possible values: \verb|true| or \verb|false|. Default is \verb|true|. Command line equivalent: \verb|-s|.
Example:
\begin{lstlisting}[breaklines,language=tex]
dirac:> #set apply_symmetry false
dirac:> \gamma_\mu\gamma_\nu
\eta_{\nu\mu} + \left[ - \frac{I}{2}\eta_{\mu\omega_{1}}\eta_{\nu\omega_{2}} + \frac{I}{2}\eta_{\mu\omega_{2}}\eta_{\nu\omega_{1}}\right]\sigma^{\omega_{1}\omega_{2}}
dirac:> #set apply_symmetry true
dirac:> \gamma_\mu\gamma_\nu
\eta_{\nu\mu}   - I\eta_{\mu\omega_{1}}\eta_{\nu\omega_{2}}\sigma^{\omega_{1}\omega_{2}}
\end{lstlisting}
or
\begin{lstlisting}
./dirac -s false
\end{lstlisting}

\subsection{Math expressions}
All input lines that are neither quit-expressions nor set-expressions are considered computable math. 
The dirac application tries to parse and compute them.
Math syntax is \LaTeX{}-like with some differences.
A valid expression consists of
\begin{itemize}
\item literals: alphanumeric sequences preceded by \verb|\|, e.g. \verb|\gamma|;
\item integer or floating-point numbers;
\item arithmetic operators \verb|+,-,*,/|;
\item subscript \verb|_|;
\item superscript \verb|^|;
\item brackets \verb|{...}| (round or square brackets are not recognized for the sake of implementation simplicity).
\end{itemize}
Subscript \verb|[head]_[tail]| and superscript \verb|[head]^[tail]| are tensorial expressions. 
Unlike \LaTeX{}, multiple non-bracketed subscripts and superscripts to a single head are possible,
but multi-level are not: \verb|\eta_\mu_\nu| is valid input while \verb|\gamma^{\eta_{\mu\nu}}| is not.

Literals' interpretation depends on their position in the input. 
Literals inside the tail of a tensorial expression are interpreted verbatim as tensor index labels.
Otherwise, only a limited number of special literals is recognized:
\begin{itemize}
\item \verb|\I| - imaginary unit;
\item \verb|\gamma| - Dirac gamma-matrix;
\item \verb|\sigma| - Dirac sigma-matrix;
\item \verb|\gamma5| - $\gamma^5$ matrix;
\item \verb|\eta| - Minkowski metric;
\item \verb|\delta| - Kronecker delta;
\item \verb|\epsilon| - Levi-Civita symbol.
\end{itemize}

The app does not perform any validation of tensorial expression consistency 
save for checking that all basic tensors have correct index counts at computation stage. 
Nevertheless, mathematically valid input results in mathematically valid output by design of the pseudo-matrix representation.

Some examples:
\begin{lstlisting}[language=tex, breaklines]
dirac:> \gamma_\lambda\gamma_\mu\gamma_\nu\gamma^\lambda
4\eta_{\mu\nu}
dirac:> \gamma_\kappa\gamma_\lambda\gamma_\mu\gamma_\nu\gamma^\kappa
\left[-2\eta_{\mu\lambda}\eta_{\omega_{1}\nu}  -2\eta_{\nu\mu}\eta_{\omega_{1}\lambda} + 2\eta_{\nu\lambda}\eta_{\omega_{1}\mu}\right]\gamma^{\omega_{1}} + 2I\epsilon_{\nu\lambda\mu\omega_{1}}\gamma^5\gamma^{\omega_{1}}
dirac:> \gamma_\mu\gamma_\nu - \gamma_\nu\gamma_\mu
 - 2I\eta_{\mu\omega_{1}}\eta_{\nu\omega_{2}}\sigma^{\omega_{1}\omega_{2}}
\end{lstlisting}

\subsection{Dirac as a library}

It is also possible to use \textit{dirac} as a library. Reusable C++ classes and routines are defined in the \verb|src/algebra| directory.
\verb|CMakeLists.txt| defines a library target \textit{dirac\_common} that includes those components. 
Top-level constructs are defined in \verb|src/algebra/Gamma.hpp|. These are
\begin{itemize}
  \item \verb|GammaPolynomial| - gamma-ring element i.e. a polynomial of Dirac matrices and Lorentz-invariant symbols with complex coefficients;
  \item \verb|CanonicalExpr| - canonical gamma-expression i.e. a linear combination of $ 1, \gamma^\mu, \sigma^{\mu\nu},
 \gamma^5\gamma^\mu, \gamma^5 $ with Lorentz-invariant coefficients;
  \item \verb|reduceGamma| - main simplification routine that converts a \verb|GammaPolynomial| to a \verb|CanonicalExpr|.
\end{itemize}

In the \verb|examples/fierz_gen| directory \textit{dirac\_common} library is applied to the problem of computing sixth-order Fierz identities stated in the introduction.

\section{Conclusions} \label{conclusions}

The \textit{dirac} software is a fully functional command line calculator for $\gamma$-matrix
polynomials. Scripting can be used to process multiple expressions in a batch ang generate \LaTeX{} documents with the output.

\bibliographystyle{elsarticle-harv}
\bibliography{refs} 

\begin{thebibliography}{11}
\expandafter\ifx\csname natexlab\endcsname\relax\def\natexlab#1{#1}\fi
\providecommand{\url}[1]{\texttt{#1}}
\providecommand{\href}[2]{#2}
\providecommand{\path}[1]{#1}
\providecommand{\DOIprefix}{doi:}
\providecommand{\ArXivprefix}{arXiv:}
\providecommand{\URLprefix}{URL: }
\providecommand{\Pubmedprefix}{pmid:}
\providecommand{\doi}[1]{\href{http://dx.doi.org/#1}{\path{#1}}}
\providecommand{\Pubmed}[1]{\href{pmid:#1}{\path{#1}}}
\providecommand{\bibinfo}[2]{#2}
\ifx\xfnm\relax \def\xfnm[#1]{\unskip,\space#1}\fi
\bibitem[{Ebert and Reinhardt(1978)}]{Ebert1978}
\bibinfo{author}{Ebert, D.}, \bibinfo{author}{Reinhardt, H.},
  \bibinfo{year}{1978}.
\newblock \bibinfo{title}{Functional approach to nuclear field theory: A
  schematic model with pairing and particle-hole forces}.
\newblock \bibinfo{journal}{Nuclear Physics A} \bibinfo{volume}{298},
  \bibinfo{pages}{60--76}.
\newblock \URLprefix
  \url{https://www.sciencedirect.com/science/article/pii/0375947478900076},
  \DOIprefix\doi{https://doi.org/10.1016/0375-9474(78)90007-6}.
\bibitem[{Fierz(1937)}]{Fierz1937}
\bibinfo{author}{Fierz, M.}, \bibinfo{year}{1937}.
\newblock \bibinfo{title}{Zur fermischen theorie des $\beta$-zerfalls}.
\newblock \bibinfo{journal}{Zeitschrift f{\"u}r Physik} \bibinfo{volume}{104},
  \bibinfo{pages}{553--565}.
\bibitem[{Itzykson and Zuber(2012)}]{itzykson2012}
\bibinfo{author}{Itzykson, C.}, \bibinfo{author}{Zuber, J.},
  \bibinfo{year}{2012}.
\newblock \bibinfo{title}{Quantum Field Theory}.
\newblock Dover Books on Physics, \bibinfo{publisher}{Dover Publications}.
\bibitem[{Kikkawa(1976)}]{Kikkawa1974}
\bibinfo{author}{Kikkawa, K.}, \bibinfo{year}{1976}.
\newblock \bibinfo{title}{Quantum corrections in superconductor models}.
\newblock \bibinfo{journal}{Progress of Theoretical Physics}
  \bibinfo{volume}{56}, \bibinfo{pages}{947--955}.
\newblock \URLprefix \url{https://doi.org/10.1143/PTP.56.947},
  \DOIprefix\doi{10.1143/PTP.56.947},
  \href{http://arxiv.org/abs/https://academic.oup.com/ptp/article-pdf/56/3/947/5455249/56-3-947.pdf}{{\tt
  arXiv:https://academic.oup.com/ptp/article-pdf/56/3/947/5455249/56-3-947.pdf}}.
\bibitem[{Kraft and Procesi(1996)}]{Kraft1996}
\bibinfo{author}{Kraft, H.}, \bibinfo{author}{Procesi, C.},
  \bibinfo{year}{1996}.
\newblock \bibinfo{title}{A Primer of Classical Invariant Theory: Preliminary
  Version}.
\newblock \bibinfo{note}{Preprint}.
\bibitem[{Kutnii(2023a)}]{DiracGithub}
\bibinfo{author}{Kutnii, S.}, \bibinfo{year}{2023}a.
\newblock \URLprefix \url{https://github.com/skutnii/dirac}.
\bibitem[{Kutnii(2023b)}]{kutnii2023bosonization}
\bibinfo{author}{Kutnii, S.}, \bibinfo{year}{2023}b.
\newblock \bibinfo{title}{Bosonization, effective action, and {R}-operation in
  a generalized {N}ambu-{J}ona-{L}asinio model}.
\newblock \href{http://arxiv.org/abs/2304.07118}{{\tt arXiv:2304.07118}}.
\bibitem[{Pal(2007)}]{PBPal2007}
\bibinfo{author}{Pal, P.B.}, \bibinfo{year}{2007}.
\newblock \bibinfo{title}{Representation-independent manipulations with dirac
  matrices and spinors}.
\newblock \URLprefix \url{https://arxiv.org/abs/physics/0703214},
  \DOIprefix\doi{10.48550/ARXIV.PHYSICS/0703214}.
\bibitem[{Peeters(2007)}]{KPeeters2007}
\bibinfo{author}{Peeters, K.}, \bibinfo{year}{2007}.
\newblock \bibinfo{title}{Cadabra: a field-theory motivated symbolic computer
  algebra system}.
\newblock \bibinfo{journal}{Computer Physics Communications}
  \bibinfo{volume}{176}, \bibinfo{pages}{550--558}.
\newblock \URLprefix
  \url{https://www.sciencedirect.com/science/article/pii/S0010465507000318},
  \DOIprefix\doi{https://doi.org/10.1016/j.cpc.2007.01.003}.
\bibitem[{Reinhardt(1978)}]{Reinhardt1978}
\bibinfo{author}{Reinhardt, H.}, \bibinfo{year}{1978}.
\newblock \bibinfo{title}{Nuclear field theory}.
\newblock \bibinfo{journal}{Nuclear Physics A} \bibinfo{volume}{298},
  \bibinfo{pages}{77--92}.
\newblock \URLprefix
  \url{https://www.sciencedirect.com/science/article/pii/0375947478900088},
  \DOIprefix\doi{https://doi.org/10.1016/0375-9474(78)90008-8}.
\bibitem[{Weyl(1946)}]{Weyl1946}
\bibinfo{author}{Weyl, H.}, \bibinfo{year}{1946}.
\newblock \bibinfo{title}{The classical groups: their invariants and
  representations}.
\newblock \bibinfo{number}{1}, \bibinfo{publisher}{Princeton university press}.

\end{thebibliography}

\end{document}